# CHAPTER 7

**Thermal Conductivity of Diamond Nanothread**

Haifei Zhan and Yuantong Gu*

*School of Chemistry, Physics and Mechanical Engineering, Queensland University of Technology (QUT), Brisbane, QLD 4001, Australia*
*Corresponding author: yuantong.gu@qut.edu.au*

**Abstract**

This chapter introduces the thermal conductivity of a novel one-dimensional carbon nanostructure – diamond nanothread. It starts by introducing the family of the diamond nanothread as acquired from density functional theory calculations and also its successful experimental synthesisation. It then briefs the mechanical properties of the diamond nanothreads as a fundamental for their engineering applications. After that, it focuses on the thermal transport properties of the diamond nanothreads by examining the influences from various parameters such as size, geometry, and temperature. Then, the application of diamond nanothread as reinforcements for nanocomposites is discussed. By the end of the chapter, future directions and their potential applications are discussed.

## XX.1 Introduction

Past decades have witnessed huge interests from both scientific and engineering communities in the one-dimensional (1D) carbon-based nanostructures, such as $sp^3$ diamond nanowire [1], $sp^2$ carbon nanotube (CNT) [2], and carbyne [3, 4]. Their excellent chemical and physical properties have enabled them as versatile and excellent integral parts for the next generation of devices [5, 6] or multifunctional materials [7]. For the $sp^2$ CNTs, their unprecedented properties have made them ideal building blocks for composite materials, nano-fibers/yarns [8-10], thin films and nanoelectromechanical systems. Researchers reported that CNT-based mechanical resonators have a quality factor as high as 5 million [5]. Diverse commercial products incorporating bulk CNT powders ranging from rechargeable batteries, automotive parts and sporting goods have already emerged [11]. In the other hand, the $sp^3$ bonded diamond nanowires, which possess unique features, such as negative electron affinity, chemical inertness, good biocompatibility, are also receiving a continuing research focus [12]. Extensive applications have been proposed for diamond nanowires, including energy absorbing material under UV laser irradiation [13], high efficiency single-photon emitters (with stable and room-temperature operation) [14], and DNA sensing [15, 16].

The attractive usages have motived researchers to seek effective ways to fabricate/synthesis different 1D carbon nanostructures [17, 18], especially the ultra-thin thread-like $sp^3$ C-H polymers. Very recently, a novel 1D $sp^3$ diamond nanostructure is synthesised from slow decompression of crystalline benzene, termed as diamond nanothread (DNT) [19]. Benefiting from their different structures, various promising applications have been expected. In this chapter, we will first introduce the big family of diamond nanothreads as evidenced from first principle calculation, and then introduce the experimental success for one kind of the diamond nanothreads. A concise introduction of their intriguing mechanical properties will be presented. Afterwards, their thermal conductivity as acquired from atomistic simulations will be discussed, followed by the discussion of their applications as reinforcements in nanocomposites. Some concludes and future directions will be given in the end of this chapter.

**XX.2 Different Diamond Nanothreads and the Synthesisation**

**XX.2.1** *The diamond nanothread family*

The diamond nanothreads (DNTs) are obtained from the compression of crystalline benzene. Before the recent experimental success, theorists predicted at least three structures of nanothread from different perspectives, including the very narrow tube (3,0) [20], polymer I (through direct simulation of high-pressure benzene) [21], and polytwistance (through the extension of twistance molecules into an extended helix) [22, 23]. Actually, the precise atomic structure of the carbon nanothreads is not yet known from experiments. In this regard, Crespi and his colleagues have conducted a systematic topological and structural enumeration of all possible bonding geometries within a one-dimensional stack of six-fold rings [21, 24] based on density functional theory (DFT) calculations. Overall, 50 distinct polymerization pathways have been identified leading to 50 distinct structures. Each topological unit cell of one or two benzene rings contains at least two bonds interconnecting each adjacent pair of rings. Specifically, 15 of these structures are identified as the most stable members, with the per carbon atom energy within 80 meV.

As illustrated in Figure 1, these 15 stable nanothreads can be divided into three groups based on their structural properties, including achiral, stiff chiral and soft chiral. All three previously proposed $sp^3$ nanothreads, i.e., tube (3,0) [20], polymer I [21], and polytwistance [22, 23], show low energy and are included in these 15 stable nanothreads.

The identification number for each model represents its bonding topology [24]. Specifically, in the achiral group, structures 1̲35̲462 and 1̲53̲624 have crystallographic units cells with 24 carbon atoms. The chiral nanothreads have long translational unit cells, but only 2 carbon atoms (for structure 1̲43̲652, the polytwistance), 6 carbon atoms for (structure 1̲35̲462), or 12 carbon atoms in the helical repeat unit. All models listed in Figure 1 have binding energies well within the range of known $C_xH_x$ hydrocarbons.

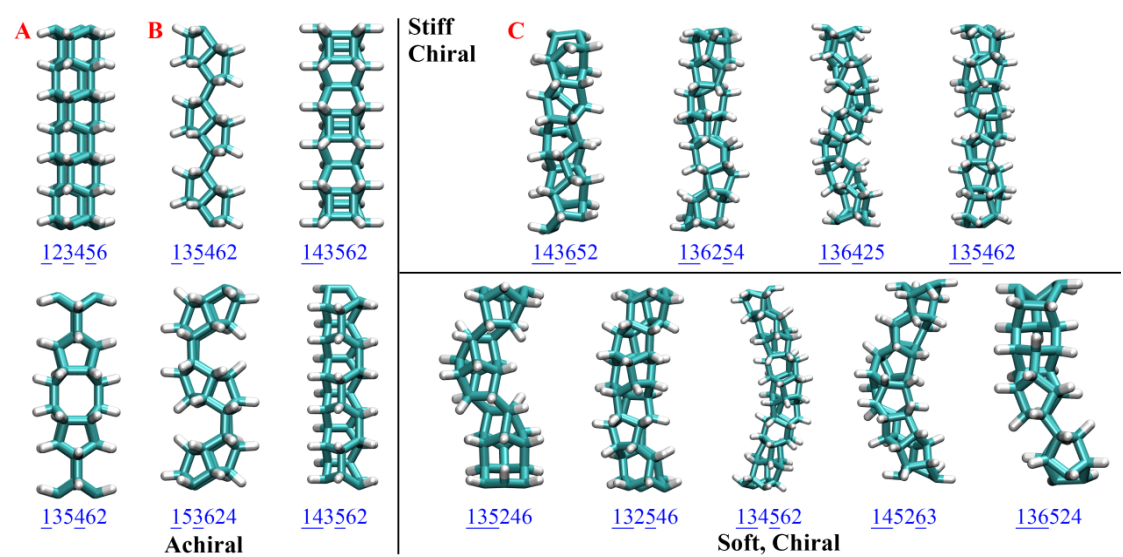

**Figure 0.1** The atomic configurations of stable nanothreads predicted from DFT calculations. A, B and C are the tube (3,0), polymer I and polytwistance, respectively. The numbers for each model represent how it is formed from the benzenes. Adapted with permission from [24]. Copyright 2016 American Chemical Society.

**XX.2.2** *Experimental synthesisation*

The experimentally synthesised diamond nanothreads are analogue to polymer I, which is a combination of tube (3,0) (structure 1̲23̲456 in Figure 1) and the so-called Stone-Wales (SW) transformation defects (structure 1̲35̲462 in Figure 1). As illustrated in Figure 2, the diamond nanothread can be regarded as hydrogenated (3,0) CNTs connected with SW transformation defects [20]. Apparently, the existence of SW transformation defects interrupts the central hollow of the structure, and differ them from the hydrogenated (3,0) CNTs.

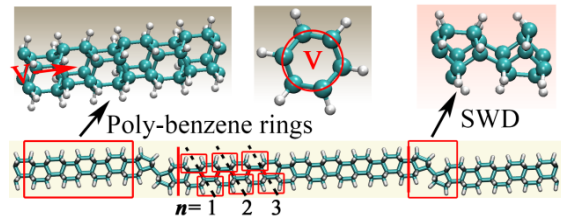

**Figure 0.2** Schematic view of the DNT as synthesised from experiments. (a) A segment of the DNT, insets show the structural representation of the poly-benzene rings and the Stone-Wales transformation defects (SWD). Reprinted from [25].

The DNT is obtained from high-pressure solid-state reaction of benzene, a complete synthesisation and characterization process can be find from Reference [19]. In general, four possible hydrogen-carbon isotopic combinations of benzene were firstly loaded as solids at low temperature into encapsulated metal gaskets to avoid evaporation. The samples were then compressed to 20 GPa at room temperature, and maintained at this pressure for one hour. Afterwards, the sample was slowly decompressed at average rate varying from 7 GPa/h to 2 GPa/h to ambient pressure. While to off-while/yellow solids were obtained from the encapsulated gaskets, which contain the DNTs as verified from a serial of characterization results.

## XX.3 Mechanical Properties

Understanding of the mechanical properties of the material is usually a pre-requisite for its engineering implementations. An initial glimpse of the mechanical properties of new nanomaterials can be achieved through DFT calculations. In this regard, DFT calculations show a wide range of Young's modulus for the DNTs, ranging from 0.08 to 1.16 TPa [24]. A more in-depth understanding of the mechanical properties of nanomaterials can be obtained from large-scale molecular dynamics (MD) simulations. This section will brief the excellent and tunable mechanical properties of DNTs as acquired from MD simulations.

### XX.3.1 *Excellent mechanical properties*

The initially examined DNTs have the same atomic structure as proposed from the experimental work [26], i.e., the DNT is comprised of by (3,0) tube and SW transformation defects. For notation convenience, a DNT unit cell with *n* poly-benzene rings between two adjacent SW transformation defects is denoted by DNT-*n*, e.g., DNT-8 has four poly-benzene rings with a length approximating 4 nm (Figure 2). In addition, the whole DNT is approximated as a solid cylinder with a diameter of 0.5 nm (the approximate distance between exterior surface hydrogens) [26].

According to the tensile simulations [26] based on ReaxFF potential [27], the DNT-8 has a tensile stiffness around 850 GPa, and the yield strain is approximately 14.9%. From the stress-strain curve, the ultimate stress of DNT-8 is about 134.3 GPa or 26.4 nN, which is over twice the strength of carbyne [4]. By introducing a virtual sticky surface to bend the DNT, a bending stiffness around 770 kcal/mol·Å is estimated. Comparing with the (5,5)CNT (rigidity on the order of 100000 kcal/mol·Å) and carbyne (rigidity around 30 to 80 kcal/mol·Å), the DNT is more like a rigid yet flexible molecule [26].

**XX.3.2** *Brittle-to-ductile transition*

Besides the excellent mechanical properties, the DNT is also exhibiting a tunable mechanical property. Reconsider the atomic structures illustrated in Figure 2, different number of poly-benzene rings can be introduced between the two adjacent SW transformation defects. It is found that the DNT unit cell has less energy when more poly-benzene rings are presented [28] (see Figure 3a). This result indicates that the DNT structure can be tailored through the poly-benzene rings. Our recent works [28] (based on AIREBO potential [29]) show that DNTs made from different unit cells (i.e., containing different number of poly-benzene rings) exhibit similar yield strength, but different yield strain and effective Young's modulus (or tensile stiffness). These results signify a strong correlation between the mechanical properties of the DNT and its geometrical structure. It is further found that the failure behaviour of the DNT is controlled by the SW transformation defect region. As illustrated in Figure 3b and 3c, there is a clear stress concentration around the defect region during tensile deformation, which is also the location that failure initiates.

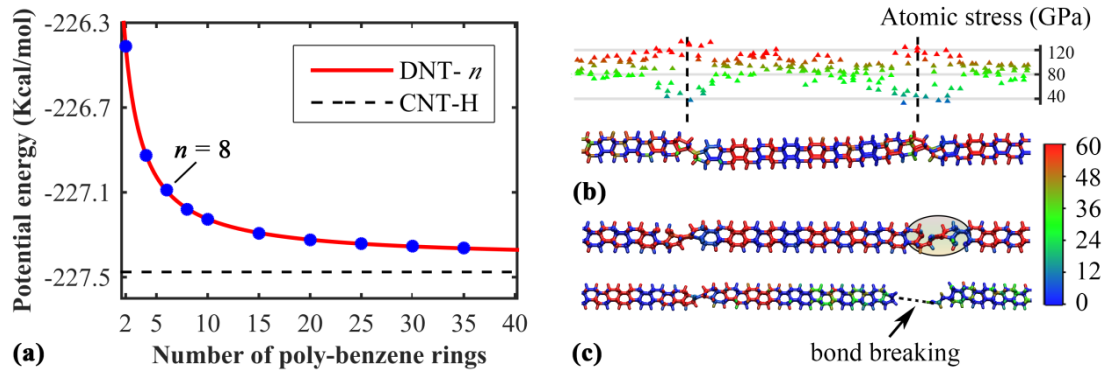

**Figure 0.3** (a) The potential energy per carbon atom for different DNT units, CNT-H represents the hydrogenated (3,0) CNT. Reprinted from [28], with permission from Elsevier. (b) The virial atomic stress distribution along the length direction at the strain of 4.6%, which clearly shows the stress concentration at the SW defect region (only carbon atom stress is presented). (c) The bond breaking configuration at the pentagon (upper, strain of 11.2%), which eventually initiates the failure of the DNT from the SW defect (lower, strain of 13.4%). Reprinted from [25].

Most strikingly, the DNT exhibits a transition from brittle to ductile characteristic. As shown in Figure 4a, the DNT-$n$ with longer poly-benzene (larger $n$) exhibits a classical brittle behaviour with a monotonically increased stress-strain curve; whereas, the DNTs with short poly-benzene, such as DNT-2, shows a clear hardening process besides the monotonically increased portion. The most interesting feature is that the hardening process has greatly deferred the failure of the DNT, and the hardening duration increases gradually with the decrease of the constituent poly-benzene length. Through the tensile deformation in a confined region, the poly-benzene rings are found to exhibit a classical brittle behavior (curve P-20), which is not affected by increasing the region length/scope (curve P-182 in Figure 4b), signifying a brittle characteristic of the poly-benzene sections. However, for the SW transformation defect region, an extra hardening process is observed (black stress-strain curve 5 in Figure 4b), which endows it with a yield strain approaching 25%, more than twice of that extracted from the confined region with only poly-benzene rings. Such results imply the ductile characteristic of the SW defect region, which is resulted from the initial bond breaking at the pentagon carbon rings. In other words, the ductility of DNT can be controlled by altering the number of SW transformation defects.

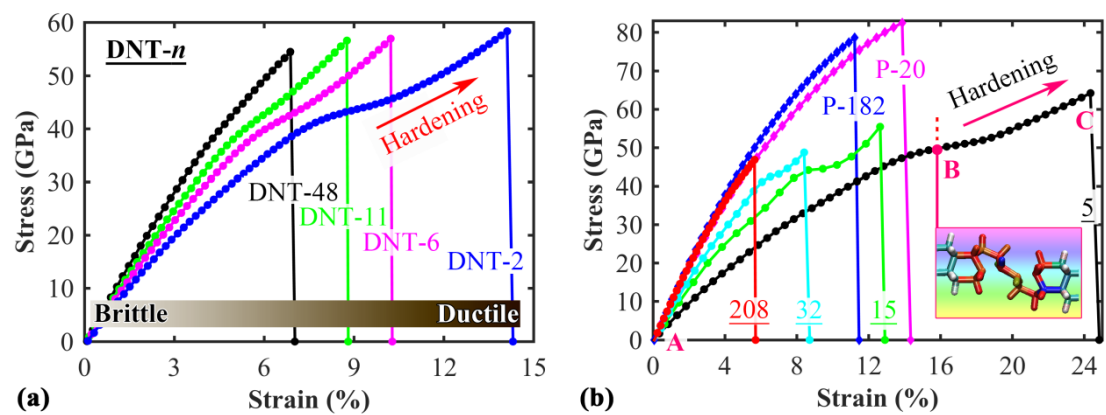

**Figure 0.4** (a) Stress-strain curves from DNTs comprised of difference constituent units with a uniform sample size of ~ 42 nm. (b) Stress-strain curves obtained from different confined regions with poly-benzene rings and SW transformation defects. P-20 and P-182 denote the two confined regions containing only poly-benzene rings with a length of 20 and 182 Å. The other underlined numbers represent the approximate length of the confined region with one SW transformation defect. Inset shows the atomic configurations of the SW defects at the strain of 16.1%. Adapted from [25].

### XX.3.3 *General mechanical properties*

Considering the diversity geometries of DNTs, it is of great interest to know how other DNTs would behave under tensile loading. As compared in Figure 5a, DNTs behaves differently from each other [30]. In general, the DNT-I (polymer I) and DNT-III (an initially helical DNT, 134562) have similar failure strain, while the DNT-II (polytwistane) shows a much larger failure strain. Such different tensile properties are originated from the different stress distributions of the DNTs under tensile loading. According to the atomic configurations, the DNT-I shows a stress concentration at the connecting carbon bonds between the two coupled pentagonal carbon rings (Figure 5b). Such stress concentration regions uniformly occur along the length direction, and the failure of the DNT is initiated from these regions with increasing strain. In comparison, the polytwistane (DNT-II) shows a generally even stress distribution pattern during tensile deformation (Figure 5c), which is reasonable as it has a uniform structure with its carbon skeleton analogous to (2,1) carbon nanotube [31]. The most striking feature is that the helical DNT-III shows a double-helix stress distribution pattern as plotted in Figure 5d. Specifically, one of the two carbon helixes is experiencing a strong tensile stress (i.e., absorbing most of the tensile strain), with the other one under minor compressive stress state. With increasing strain, failure is triggered along the helix with tensile stress. Overall, these results suggest that there is big room to tune the mechanical properties of DNTs.

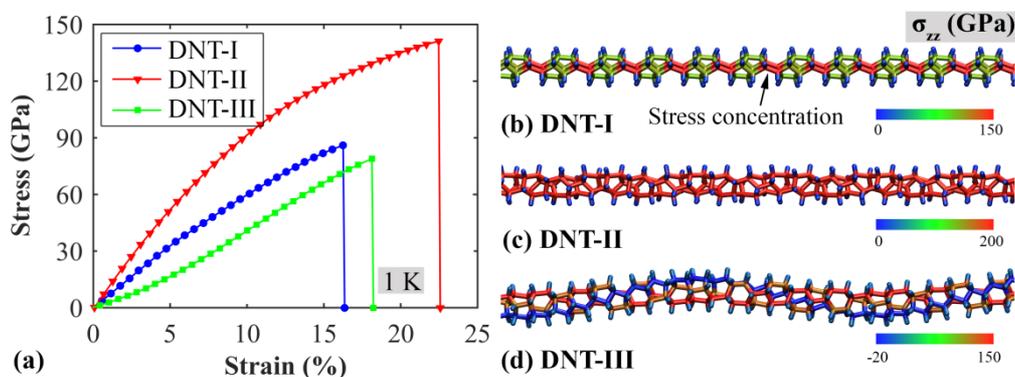

**Figure 0.5** (a) Comparisons of the stress-strain curves of the three DNTs at 1 K; (b-d) The atomic stress distribution of the three DNTs at the strain of 12% (before failure) along the stretch direction. Atoms are coloured according to the atomic stress along the length direction. Adapted from [30], with permission from Elsevier.

Since the DNT is normally comprised of by pentagon, hexagon, heptagon and octagon carbon rings, its carbon bond length ranges from 1.51 to 1.67 Å (at 0 K) [24]. Owing to these long and non-uniformly distributed carbon bonds, the temperature is found to exert an obvious influence to their tensile properties [30]. Figure 6a shows the failure strength of DNT-III decreases from ~ 79 GPa to ~ 27 GPa when the temperature changes from 1 K to 300 K, corresponding to more than 60% reduction. Similar results are also observed from DNT-I and DNT-II. As compared in Figure 6b, the relative failure strength ($\sigma_{fr}$) decreases with the increase of temperature for all examined DNTs. Here $\sigma_{fr} = \sigma_{fN} / \sigma_{f0}$, with $\sigma_{fN}$ and $\sigma_{f0}$ represent the failure strength at the temperature of $N$ and 1 K, respectively. In comparison, the yield strain shows a similar decreasing tendency, but Young's modulus appears independent of temperature.

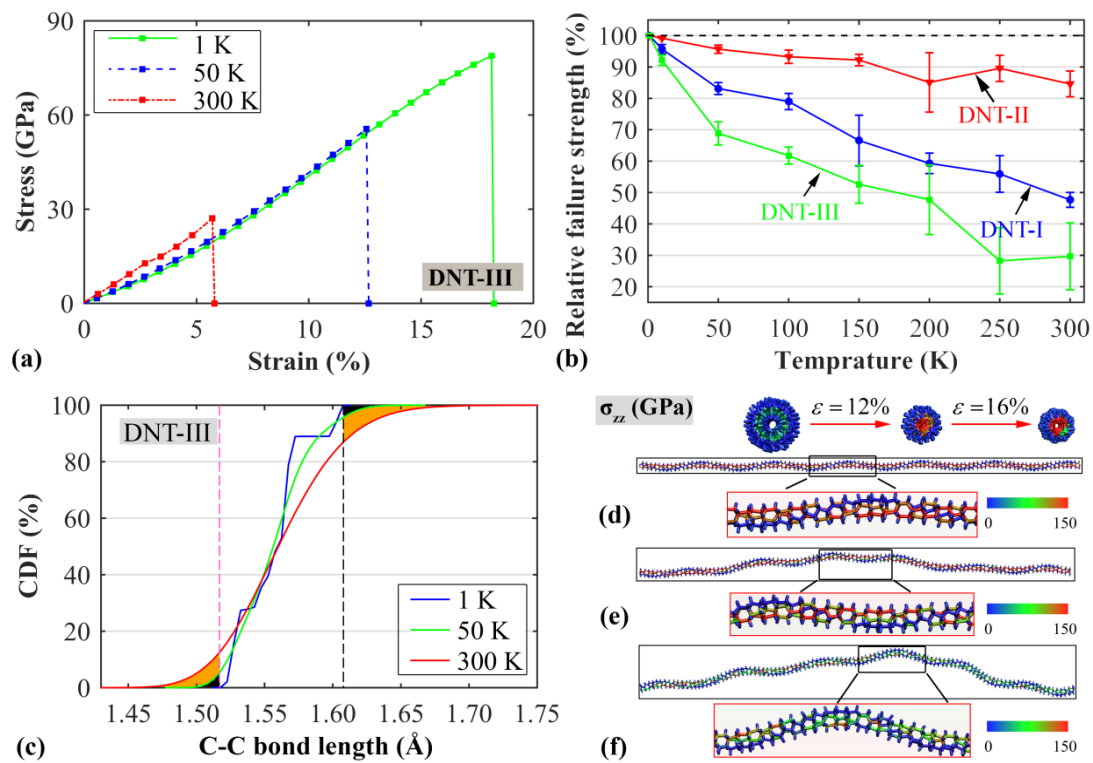

**Figure 0.6** (a) Comparisons of the stress-strain curves at the temperature of 1, 50 and 300 K for DNT-III. (b) The relative failure strength of the three DNTs as a function of temperature. Errorbar represents the relative standard deviation of the failure strength calculated from four different simulations with relaxation time ranging from 1 to 2.5 ns. (c) The cumulative density function (CDF) of the carbon bond length of DNT-III at the temperature of 1, 50 and 300 K. The left and right dash lines indicate the shortest and longest carbon bond length of DNT-III at 1 K, respectively. (d) The atomic stress distribution of DNT -III at the strain of 12% along the stretch direction under 1 K.

Upper shows the shrinkage of the DNT's cross-section with increasing strain; The atomic stress distribution of DNT -III at the strain of: (e) 11% under 50 K and (f) 4% under 300 K, along the stretch direction. Atoms are coloured according to the atomic stress along the length direction.

The significant temperature impacts on the mechanical properties of DNTs can be explained from the perspective of the coefficient of thermal expansion (CTE) [32]. Figure 6c shows the cumulative density function (CDF) of the carbon bond length of DNT-III at the temperature of 1, 50 and 300 K. As is seen the carbon bond length has a larger range at higher temperature, indicating both bond shortening and lengthening at increased temperatures. The presence of longer carbon bonds, with correspondingly lower C-C bond strength, will lead to bond failure at lower stress, and thus lead to smaller failure strength/strain. In addition, the free lateral/bending vibration of the DNT at higher temperature is also a crucial factor that affects its mechanical performance. As compared in Figure 6d, the DNT-III can maintain its helical structure well during tensile deformation at low temperature (1 K). However, at higher temperature (50 K in Figure 6e), obvious offset of the DNT's axis is observed, which is resulted from the lateral vibration. These offset is more obvious in the case of 300 K (Figure 6f), which changed the double-helix stress distribution pattern as observed at 1 K. In all, the temperature exerts a significant impact on the mechanical properties of the DNT.

## XX.4 Thermal Conductivity

Accompanying with the excellent and tailorable mechanical properties, the DNT is also showing intriguing thermal transport properties. This section will introduce the thermal transport properties of DNTs with SW transformation defects as obtained from non-equilibrium molecular dynamics (NEMD) simulations (based on AIREBO potential [29]).

### XX.4.1 *Superlattice thermal transport characteristic*

As aforementioned, the SW transformation defects interrupt the continuity of the (3,0) tube, which also induce the "interfacial thermal resistance" or Kapitza resistance (KR). As plotted in Figure 7a, there exists a clear temperature jump (around 4.2 K) at the region with SW transformation defect. Different from the traditional circumstance that the KR is triggered when the heat flows across the contacting interface between different materials, the observed KR occurs between same poly-benzene rings. By approximating the temperature gradient along the heat flux direction as $\Delta T / L$, a low thermal conductivity of about 35.6 ± 4.7 W/mK is estimated for the DNT-55 (with one SW transformation defects). Here, $\Delta T = T_h - T_c$ is the temperature difference between the heat source $T_h$ and heat sink $T_c$, and $L$ is the length of DNT. Compared with the ultra-thin (2,1) CNTs [33], and also the (3,0) CNT (63.1 ± 6.3 W/mK), the DNT shows a small thermal conductivity. From Figure 7b, the DNT-8 shows significant reduction in phonon lifetimes almost over the entire range of frequency comparing with that of the pristine (3,0) CNT. This result indicates additional phonon scattering at the SW transformation defect region, which thus leads to low thermal conductivity of DNTs.

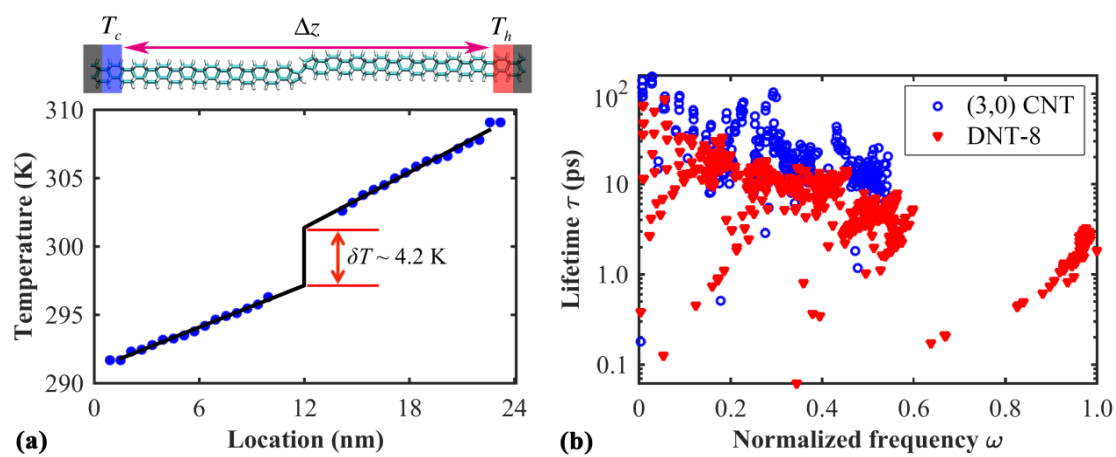

**Figure 0.7** (a) Temperature profile of the DNT-55 (length of ~ 24 nm), which contains only a single SW transformation. The atomic configuration is only a schematic representation of the studied model. (b) Comparisons of the DOS between CNT, the poly benzene rings portion of DNT, and the SW transformation region of DNT. The

DOS is extracted from the autocorrelation function of the atomic velocities obtained from the MD simulations [34]. The mode lifetimes as a function of the normalized frequency for (3,0) CNT and DNT-8 as calculated from the package Jazz [35, 36]. Adapted from [28], with permission from Elsevier.

Besides the abrupt temperature drop as induced by the phonon scatting at the SW transformation defect region, the DNT is also exhibiting a superlattice thermal transport characteristic [28]. As illustrated in Figure 8a, the thermal conductivity increases as the sample length increases, irrespective of the number of poly-benzene rings. In particular, for a given sample length, as the number of poly-benzene rings increases, the estimated thermal conductivity initially decreases for smaller $n$, and then undergoes a relatively smooth increase. Through the normalization of the results by the minimum thermal conductivity, a consistent scaling behaviour is observed (Figure 8b). Such interesting feature have also been observed in thin-film [37], nanowire [38] and superlattices [39, 40].

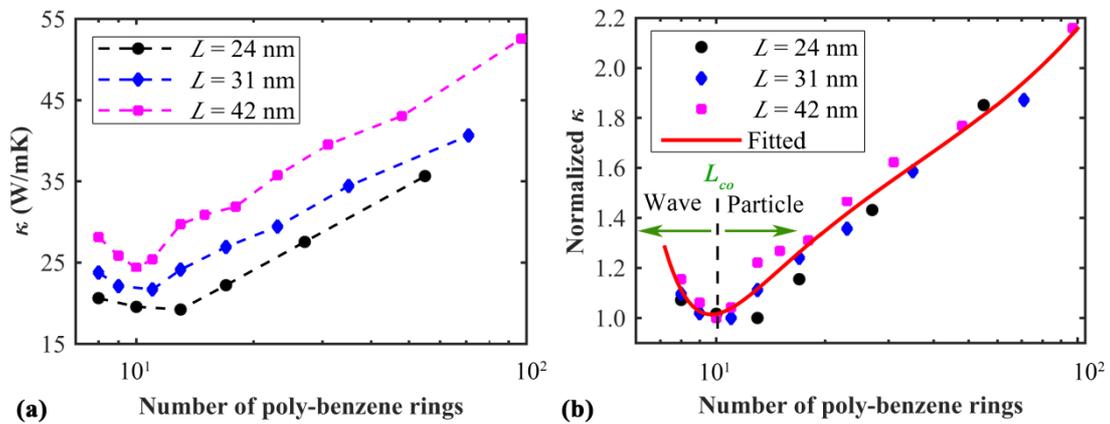

**Figure 0.8** (a) Thermal conductivity plotted as a function of DNT length and constituent unit cells; (b) The thermal conductivity normalized by its respective minima shows a general valley trend. This indicates universal critical length scales which correspond to the phonon coherence length $L_{co}$, estimated as 5.6 nm. Reprinted from [28], with permission from Elsevier.

The superlattice thermal transport characteristic can be explained from the perspective of the phonon coherence length. Based on the phonon DOS [39, 41], the phonon coherence length of DNT can be estimated from $L_{co} = \tau_{co} v_D$ with $\tau_{co}$ and $v_D$ being the coherence time and Debye velocity, respectively. The coherence time can be estimated according to $\tau_{co} = \int_0^{\omega_m} [g(\omega) D_{BE}(\omega)]^2 d\omega$. Here, $\omega_m$ is the maximum frequency of all phonon modes, and $g(\omega)$ is the normalized DOS according to the Bose-Einstein distribution $D_{BE}$ (i.e., $\int_0^{\omega_m} g(\omega) D_{BE}(\omega) d\omega = 1$). The Debye velocity is estimated from $3/v_D^3 = 1/v_L^3 + 2/v_T^3$, with $v_L$ and $v_T$ representing the sound velocity for the one longitudinal and two transverse branches, respectively [42] (which can be derived from the elastic Lame's constants $\lambda$, $\mu$, and the mass density $\rho$ [43]). Overall, the primitive estimation shows that the DNT possess a coherence length around 5.6 nm, which is close to the length of a DNT-11 unit cell and agrees well with the result in Figure 8b. Therefore, for the DNT with poly-benzene number smaller than around 10 (left region in Figure 8b), phonon transport is largely dominated by wave effects including constructive and destructive interferences arising from interfacial modulation. While, in the right region, phonon waves lose their coherence and transport is more particle-like. Similar results are also observed from graphene/h-BN superlattices [39], i.e., the minimum thermal conductivity occurs at a transition from wave-dominated to particle-dominated transport region.

## XX.4.2 *Length and temperature dependence*

Similar as observed from other 1D nanomaterials, such as diamond nanorods [44], or CNTs [33], the thermal conductivities of DNT has a strong length dependency. It is found that the thermal conductivity increases with the length of DNT. As shown in Figure 9a, the inverse of the estimated thermal conductivity for the DNTs follows a linear scaling relationship with the inverse of its length. Such linear scaling behaviour follows the linear relationship as derived from the kinetic theory [45, 46], that is, $1/\kappa = 1/\kappa_\infty (1+\lambda/L)$. Here, $\kappa$ and $\kappa_\infty$ are the size dependent and converged (when sample size is large enough) thermal conductivity, respectively. $L$ is the sample length, and $\lambda$ is the MFP. This linear scaling behaviour of thermal conductivity has also been commonly found in other 1D nanostructures [45, 47-49], 2D nanoribbons [50, 51], and bulk materials [46].

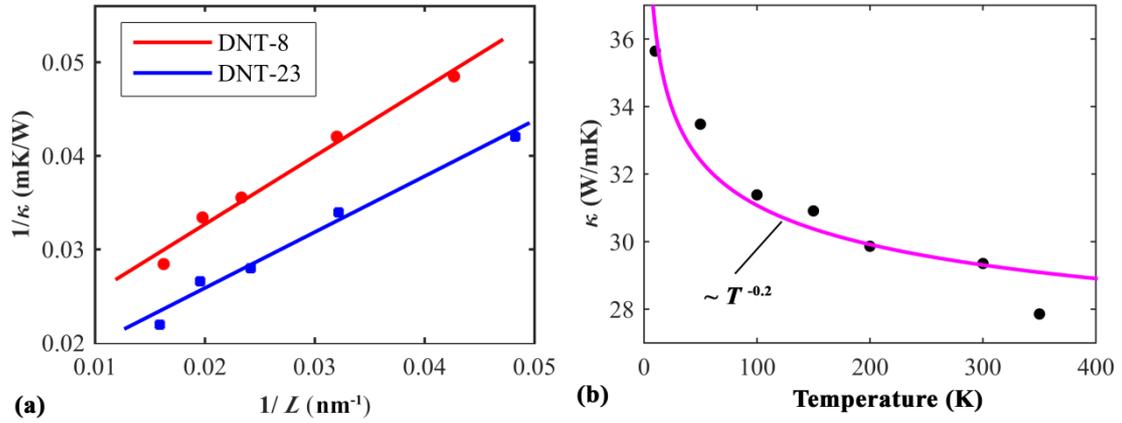

**Figure 0.9** (a) The inverse of the thermal conducted versus the inverse of sample length, which exhibits a common heuristic linear scaling relationship. (b) Thermal conductivity of DNT-13 (size of ~ 42 nm) as a function of temperature. The solid line is fitted with a power-law relation. Adapted from [28], with permission from Elsevier.

In addition, the thermal conductivity of DNTs also changes with temperature. As illustrated in Figure 9b, the thermal conductivity gradually decreases with temperature, and roughly obeys the $T^{-\alpha}$ law. Similar results are also obtained by using equilibrium molecular dynamics (EMD) simulation [52]. According to the Boltzmann transport equation, the phonon thermal conductivity is expressed as $\kappa(T) = \sum_j C_j(T) v_j^2 \tau_j(T)$ at a given temperature [53]. Here $C_j$, $v_j$ and $\tau_j$ are the specific heat, phonon group velocity and phonon relaxation time of phonon mode $j$. The phonon relaxation time is mainly limited by boundary scattering and the three-phonon umklapp scattering processes [54]. At high temperature, the role of umklapp scattering is dominating, and the scattering rate of the Umklapp process is proportional to temperature [55, 56]. That is the thermal conductivity drops steadily with the increasing temperature for homogeneous materials, and obeys a power-law dependence with an exponent of −1. The results in Figure 9b show an obvious deviation from $T^{-1}$ law at high temperature, which is considered as resulted from the weakening effect as induced by the additional phonon scattering at SW transformation defect regions.

### XX.4.3 *Comparisons with carbyne chain*

It is of great interest to compare the thermal properties with other 1D carbon chain, the carbyne chain. Essentially, carbyne chain is a 1D sp-hybridized carbon allotrope which has two forms including α-carbyne (polyyne, with alternating single and triple bonds) and β-carbyne (cumulene, with repeating double bonds). Though the existence of this ultrathin carbon has been debated [57], several experimental works have reported its synthetization or fabrication [58, 59]. Recent work shows that cumulene can transition to polyyne at the temperature of 499 K, and the transition is very fast within 150 fs [60].

Using NEMD simulation, the thermal conductivity for cumulene (with 50 carbon atoms) is about 83 W/mK at 480 K. Considering that the cumulene is metallic, the electronic contribution to the thermal conductivity can be estimated from the Wiedemann-Franz law $\kappa_e = \sigma L_w T$. Here, $L_w$ is the Lorenz number, $T$ is temperature and $\sigma$ is electrical conductivity (~ $1.6 \times 10^2$ S/m). In this regard, the electronic contribution is around $1.11 \times 10^{-3}$ W/mK, which is ignorable compared with the phonon thermal conductivity. Meanwhile, a much smaller thermal conductivity is estimated for the same polyyne chain, around 42 W/mK at 500 K. These results suggest that the carbyne chain has a same order of thermal conductivity as that estimated for DNT from both EMD simulations [52] and NEMD simulations [28]. Additionally, the simulations show that with a small percentage of defects (4%), the thermal conductivity of polyyne chain will suffer a substantial decrease (from 42 to 5.5 W/mK). Here, the defect in polyyne chain is the carbon bond that is shorter than that of the single bond but longer than double and triple bonds, which is seen when the heating rate is high than around 2.5 K/fs or if the cumulene chain is initialized at a temperature over 499 K [60] in the DFT calculations.

The underlying mechanism of the remarkable thermal transport difference among cumulene, polyyne, and polyyne with 4% defects can be explained by comparing the heat current autocorrelation function (HCACF, $C_{JJ}(t) =< J_i(0)J_i(t)>$). As is known, the decay of HCACF in bulk material is exponential, and the initial fast decay is due to high-frequency phonon modes while slow decay arises from low-frequency phonon modes [60]. According to Figure 10, the cumulene exhibits a rapid initial decay followed by a long time decay (around 15 ps), while, the polyyne shows a shorter decay time, about 10 ps. Theoretically, high-frequency phonons have a limited contribution to thermal conductivity due to their low group velocity. That is, the large difference in relaxation time for long-wavelength phonons along the chain is responsible for the remarkable difference in thermal conductivity between cumulene and polyyne.

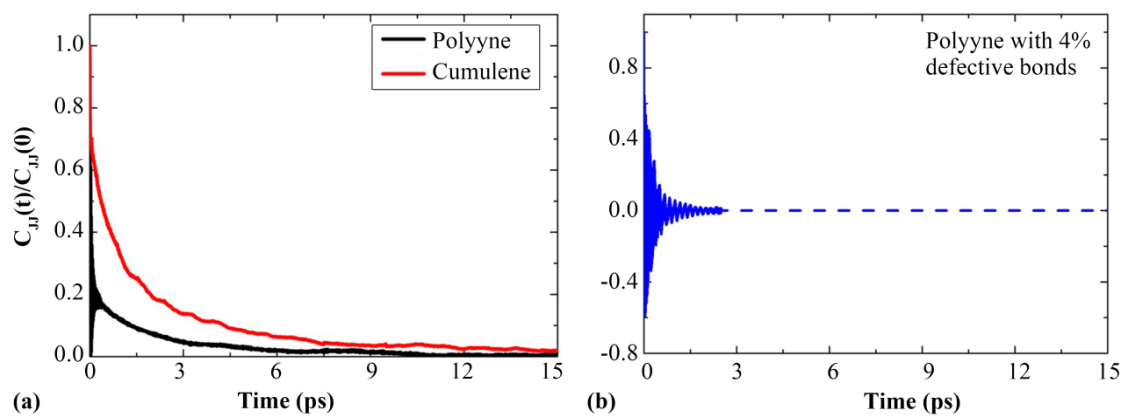

**Figure 0.10** Heat flux autocorrelation functions for carbyne chains. (a) Cumulene and perfect polyyne chains. (b) Polyyne chain with 4% defective bonds. The temperature of cumulene is 480 K, and the temperature of polyyne is 500 K. Adapted with permission from [60]. Copyright 2016 American Chemical Society.

Meanwhile, the HCACF of defective polyyne is much lower than that of the pristine polyyne. As shown in Figure 10b, the HCACF of the defective polyyne has regular high-frequency oscillations, and the decay time is remarkably shorter comparing with that of the pristine polyyne and cummulene. These defective carbon bonds are supposed to induce increased phonon scattering, and thus introduces localized vibrational modes that reduces thermal conductivity. In all, both carbyne chain and DNT has a tunable thermal conductivity. The difference is that the carbyne chain is realized through the phase change, while the DNT is through the structural change.

**XX.5 Applications of Diamond Nanothread**

The intrinsic geometry of the DNTs and also their excellent mechanical properties are expected to offer great potential applications in the field of nanocomposites. In this context, the applications of DNT as reinforcements for nanocomposites have been discussed recently by choosing the high-density polyethylene (PE) as a representative polymer [61]. Figure 11a shows the polymer model comprised of by pristine linear polyethylene molecule and DNT-2, which was constructed through the packing module [62] using Materials Studio (version 6.0) from Accelrys Inc. The initial size of the polymer system is about $4 \times 4 \times 6$ nm$^3$, and the polymer has an experimentally measured density of 0.92 g/cm$^3$ [63]. The atomic interactions within the polymer and the DNT were described by the polymer consistent force field (PCFF) [64], which has been shown to reproduce well the mechanical properties, compressibility, heat capacities and cohesive energies of polymers and organic materials. The 6-9 Lennard-Jones (LJ) potential was applied to describe the van der Waals (vdW) interactions with a cutoff of 12.5 Å, and the Ewald summation method was employed to treat the long-range Coulomb interactions [65]. The pull-out of the DNT is realized by applying a constant velocity to one end of the DNT and fixed the four edges of the polymer (Figure 11b and 11d), which closely mimics the experimental setup [66, 67].

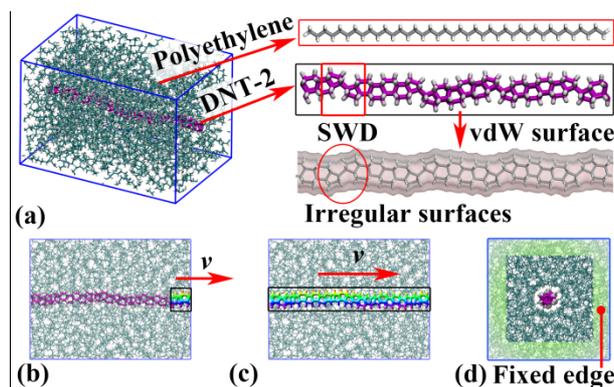

**Figure 0.11** (a) A periodic PE composite model containing DNT-2. The inset shows the atomic configurations of the polyethylene, DNT-2 and also its van der Waals surface; Schematic view of the pull-out setting with the velocity load applied to: (b) the end of the DNT, and (c) the whole rigid DNT, the loading region of the DNT is treated as a rigid body; (d) Cross-sectional view of the PE composite shows the fixed edges during pull-out simulation. Adapted from [61], with permission from John Wiley and Sons.

As illustrated in Figure 12a, the total energy change ($\Delta E_{po}$) increases with the sliding distance and saturates at a value of ~ 9.5 eV when the DNT is fully pulled-out from the PE matrix. Basically, the potential energy of the whole system comprises the sum of all the covalent bonds, van der Waals and Coulombic interactions. Considering that there is no bond breaking or disassociation under the PCFF force field, the total energy change $\Delta E_{po}$ is mainly attributed to the change of the vdW and electrostatic energy. As is seen in Figure 12a, the total vdW energy change ($\Delta E_{vdW}$) is similar to the total energy change (~ 9.5 eV), indicating the electrostatic interaction is very weak compared with the vdW interaction at the DNT/PE interface. Meanwhile, the total energy change shows significant fluctuations, which results from the local deformation or relaxation of the polymer matrix. From Figure 12a, these fluctuations are only seen in the profile of the total potential change of the polymer matrix, and the energy curve for the DNT ($\Delta E_{po}^{DNT}$) does not show significant fluctuations. It is supposed that during the pull-out, the attractive force between the embedded DNT and PE will induce local deformation to the adjacent PE chains, and the release of the pre-occupied space will allow a free relaxation of the polymer matrix that surrounded the DNT tail. These local deformation and also free relaxation as evidenced from the absolute relevant atomic displacements in the composite structure (Figure 12b) are accompanied by the absorption and release of the strain energy and thus lead to local fluctuations to the total potential energy. In addition, the further pull-out test by treating the DNT as a rigid body (Figure 10c, which has been commonly applied for CNT and polymer system [68-70]), suggests a similar total potential energy change (~ 9.5 eV). These results signify that the load transfer at the DNT/PE interface is dominated by the vdW interactions.

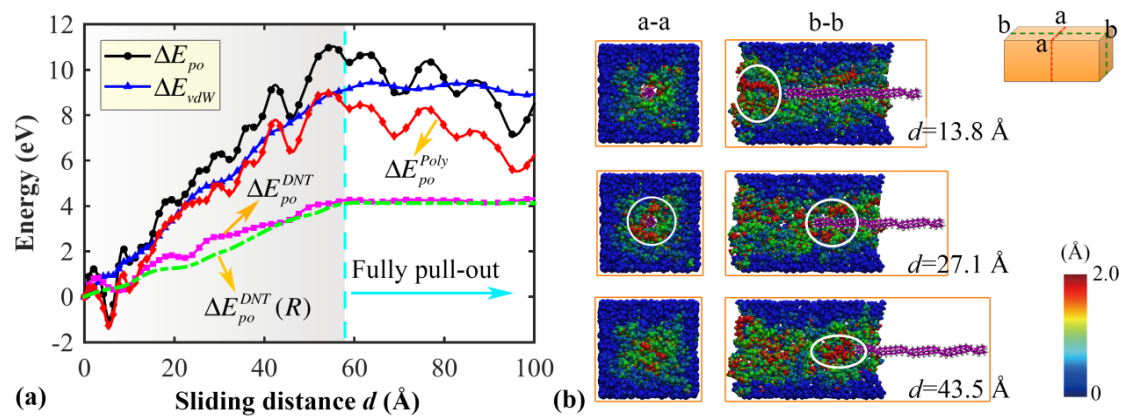

**Figure 0.12** (a) The potential energy change of the whole PE composite ($\Delta E_{po}$), the polymer matrix ($\Delta E_{po}^{Poly}$) and the DNT ($\Delta E_{po}^{DNT}$), and the total vdW energy change ($\Delta E_{vdW}$), as a function of the sliding distance. The potential energy change for the DNT while it is taken as a rigid body is also presented for comparisons ($\Delta E_{po}^{DNT}(R)$); (b)

Cross-sectional views of the polymer atomic configurations at different sliding distance. The atoms are coloured according to the absolute relevant atomic displacements calculated according the adjacent atomic configurations. Atoms are coloured as red for displacements over 2 Å. Reprinted from [61], with permission from John Wiley and Sons.

With the obtained total potential energy change ($\Delta E_{po}$) or the total vdW energy change ($\Delta E_{vdW}$), the load transfer at the composite and reinforcement interface can be quantified by estimating the interfacial shear strength (ISS, $\tau$) [70, 71] based on $\Delta E_{po} = \int_0^L F(x)dx$. Here, $x$ is the sliding distance and $L$ is the embedded length of the DNT. $F(x)$ is the shear force, which can be calculated as $F(x) = \pi D(L-x)\tau$. Thus, the interfacial shear strength can be derived as $\tau = 2\Delta E_{po} / \pi DL^2$, which yields to an ISS of ~ 58 MPa at the PE/DNT-2 interface. In comparison, the same pull-out test of an ultra-thin (4,0) CNT from the PE matrix shows a total potential energy change around 12 eV, leading to an ISS of ~ 54 MPa. In literature, the interfacial shear strength at the PE/CNT(10,10) interface [72] is about 33 MPa and PE/CNT(10,0) interface [68] is about 133 MPa. These results suggest that despite the hydrogenated surface, the DNT has comparable interfacial shear strength compared to CNT. It is supposed that during the pull-out, the PE matrix will deform in both normal and interfacial (pull-out) directions. That is, the weaker vdW interaction at the PE/DNT interface is in effect assisted by mechanical interlocking, and therefore leads to a comparable ISS with that of the PE/CNT interface.

Given the high tunability of the DNT's structure, further pull-out tests have also been conducted to explore the effective ways to enhance the load transfer at the interface. As compared in Figure 13a, the similar ISS for the polymer with different DNT structures suggest that the number of Stone-Wales defects exerts insignificant influence on ISS for the examined models. However, by increasing the polymer density from 0.92 to 1.00 g/cc [63], an obvious enhancement of ISS is observed, which is due to the increased vdW interaction sites. Furthermore, by adding 5% of randomly distributed -$C_2H_5$ functional groups to the DNT surface (as is commonly applied to CNTs [72-74]), an increase of about 20% in ISS is observed (from ~ 58 MPa to ~ 68 MPa). According to Figure 13b, the additional functional groups will not only increase the vdW interaction sites between DNT and PE matrix, but also result in more local deformation of polymer during pull-out. Apparently, the functional groups appear more effective in enhancing the load transfer at the interface comparing with the polymer density and DNT structure.

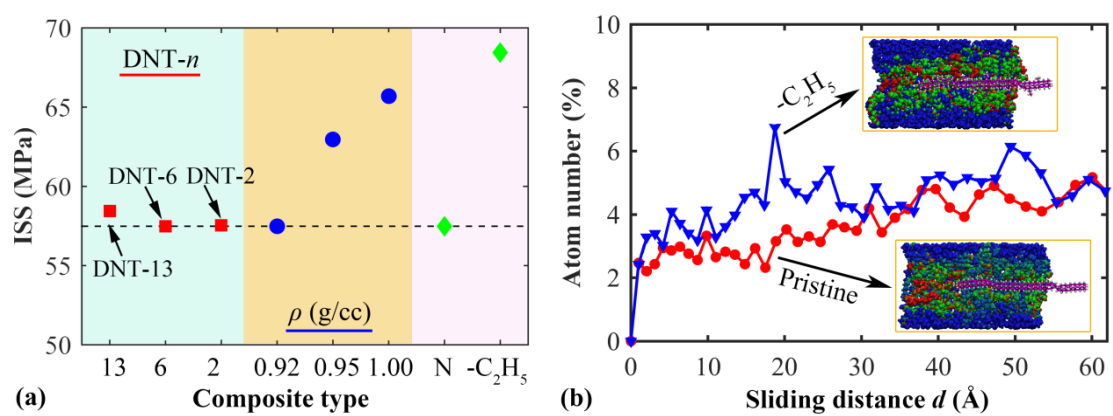

**Figure 0.13** (a) Estimated ISS for PE composites with different DNTs (DNT-13 with one SW defect, DNT-6 with two SW defects and DNT-2 with four SW defects, left panel), with different initial density (0.92, 0.95 and 1.00 g/cc, middle panel), with pristine DNT-6 (N) or –$C_2H_5$ functionalized DNT-6 (right panel). (b) The percentage of PE atoms with atomic displacements over 2 Å. Insets show the atomic configurations at the sliding distance of ~ 19 Å, and atoms with atomic displacements over 2 Å are colored red. Reprinted from [61], with permission from John Wiley and Sons.

Overall, the un-covalently embedded DNT is found to possess similar interfacial shear strength with the polymer comparing with that of CNT. Considering its hydrogen surface, it is easy to introduce cross-links between DNT and the host polymers, which will thus enable strong load transfer between DNT and polymer. Recall the excellent mechanical properties, the mechanical properties of the polymer would be greatly enhanced. Meanwhile, the relative higher thermal conductivity of DNT will also greatly enhance the thermal performance of the polymers, which has already seen in the CNT or graphene embedded nanocomposites [74-76].

**XX.6 Summary and Future Directions**

As a novel 1D ultra-thin carbon nanostructure, the research on diamond nanothread is still at the early stage. The experimental synthesisation success and also the DFT calculations have suggested that they are many different kinds of diamond nanothreads. Current MD simulations have shown that DNT possesses not only excellent and tunable mechanical properties, but also a superlattice thermal transport characteristic. Considering the diversity of their structures, a great effort is still expected in interpreting their physical, chemical, and other properties.

Along with their physical and chemical properties characterization, extensive attempts are also expected to unveil their potential applications in various fields. It is expected that this novel diamond nanothread would offer the potential for improved load transfer through covalent bonding, efficiently transferring their mechanical strength to a surrounding matrix and thus allowing for technological exploitations in fibres or fabrics [19]. In this regard, the applications of DNT as reinforcements for nanocomposites have already been discussed [61], and it is found that the un-covalently embedded DNT shows similar interfacial shear strength with the polymer comparing with that of CNT. While, considering its hydrogenated surface, it is relatively easy to introduce covalent interactions, or even cross links with other DNT threads. This could create multi-thread structures, and it is expected to greatly enhance the load transfer. Additionally, nanothreads are also expected to allow for very large electric fields at their ends and a negative electron affinity similar to that reported for hydrogen-terminated nanodiamond surfaces [77], making them attractive for field-emission applications [19].

## Acknowledgment

The studies of the mechanical and thermal properties of diamond nanothread were supported by the ARC Discovery Project (DP130102120), the Australian Endeavour Research Fellowship, and the High Performance Computer resources from the Queensland University of Technology (Australia) and Institute of High Performance Computing (Singapore).

## References

[1] Sun L, Gong J, Zhu D, Zhu Z and He S 2004 Diamond nanorods from carbon nanotubes *Adv. Mater.* **16** 1849-53
[2] Iijima S 1991 Helical microtubules of graphitic carbon *Nature* **354** 56-8
[3] Nair A, Cranford S and Buehler M 2011 The minimal nanowire: Mechanical properties of carbyne *EPL (Europhysics Letters)* **95** 16002
[4] Liu M, Artyukhov V I, Lee H, Xu F and Yakobson B I 2013 Carbyne from First Principles: Chain of C Atoms, a Nanorod or a Nanorope *ACS Nano* **7** 10075-82
[5] Moser J, Eichler A, Güttinger J, Dykman M I and Bachtold A 2014 Nanotube mechanical resonators with quality factors of up to 5 million *Nat Nano* **9** 1007-11
[6] Chaste J, Eichler A, Moser J, Ceballos G, Rurali R and Bachtold A 2012 A nanomechanical mass sensor with yoctogram resolution *Nat. Nanotechnol.* **7** 301-4
[7] Liu Z, Fang S, Moura F, Ding J, Jiang N, Di J, et al. 2015 Hierarchically buckled sheath-core fibers for superelastic electronics, sensors, and muscles *Science* **349** 400-4
[8] Jiang K, Li Q and Fan S 2002 Nanotechnology: Spinning continuous carbon nanotube yarns *Nature* **419** 801-
[9] Zhang X, Jiang K, Feng C, Liu P, Zhang L, Kong J, et al. 2006 Spinning and Processing Continuous Yarns from 4‐Inch Wafer Scale Super‐Aligned Carbon Nanotube Arrays *Adv. Mater.* **18** 1505-10
[10] Ma W, Liu L, Yang R, Zhang T, Zhang Z, Song L, et al. 2009 Monitoring a micromechanical process in macroscale carbon nanotube films and fibers *Adv. Mater.* **21** 603-8
[11] De Volder M F, Tawfick S H, Baughman R H and Hart A J 2013 Carbon nanotubes: present and future commercial applications *Science* **339** 535-9
[12] Yu Y, Wu L and Zhi J 2014 Diamond Nanowires: Fabrication, Structure, Properties, and Applications *Angew. Chem. Int. Ed.* **53** 14326-51
[13] Coffinier Y, Szunerits S, Drobecq H, Melnyk O and Boukherroub R 2012 Diamond nanowires for highly sensitive matrix-free mass spectrometry analysis of small molecules *Nanoscale* **4** 231-8
[14] Babinec T M, Hausmann B J, Khan M, Zhang Y, Maze J R, Hemmer P R, et al. 2010 A diamond nanowire single-photon source *Nat. Nanotechnol.* **5** 195-9
[15] Yang N, Uetsuka H, Osawa E and Nebel C E 2008 Vertically aligned diamond nanowires for DNA sensing *Angew. Chem. Int. Ed.* **47** 5183-5
[16] Yang N, Uetsuka H and Nebel C E 2009 Biofunctionalization of Vertically Aligned Diamond Nanowires *Adv. Funct. Mater.* **19** 887-93
[17] Zhao X, Liu Y, Inoue S, Suzuki T, Jones R and Ando Y 2004 Smallest carbon nanotube is 3 Å in diameter *Phys. Rev. Lett.* **92** 125502

[18]    Plank W, Pfeiffer R, Schaman C, Kuzmany H, Calvaresi M, Zerbetto F, et al. 2010 Electronic structure of carbon nanotubes with ultrahigh curvature *ACS Nano* **4** 4515-22
[19]    Fitzgibbons T C, Guthrie M, Xu E-s, Crespi V H, Davidowski S K, Cody G D, et al. 2015 Benzene-derived carbon nanothreads *Nat. Mater.* **14** 43-7
[20]    Stojkovic D, Zhang P and Crespi V H 2001 Smallest nanotube: Breaking the symmetry of sp 3 bonds in tubular geometries *Phys. Rev. Lett.* **87** 125502
[21]    Chen B, Hoffmann R, Ashcroft N W, Badding J, Xu E and Crespi V 2015 Linearly Polymerized Benzene Arrays As Intermediates, Tracing Pathways to Carbon Nanothreads *J. Am. Chem. Soc.* **137** 14373-86
[22]    Olbrich M, Mayer P and Trauner D 2014 A step toward polytwistane: synthesis and characterization of C 2-symmetric tritwistane *Org. Biomol. Chem.* **12** 108-12
[23]    Barua S R, Quanz H, Olbrich M, Schreiner P R, Trauner D and Allen W D 2014 Polytwistane *Chemistry-A European Journal* **20** 1638-45
[24]    Xu E-s, Lammert P E and Crespi V H 2015 Systematic enumeration of sp3 nanothreads *Nano Lett.* **15** 5124-30
[25]    Zhan H, Zhang G, Tan V B C, Cheng Y, Bell J M, Zhang Y-W, et al. 2016 From brittle to ductile: a structure dependent ductility of diamond nanothread *Nanoscale* **8** 11177-84
[26]    Roman R E, Kwan K and Cranford S W 2015 Mechanical Properties and Defect Sensitivity of Diamond Nanothreads *Nano Lett.* **15** 1585-90
[27]    Van Duin A C, Dasgupta S, Lorant F and Goddard W A 2001 ReaxFF: a reactive force field for hydrocarbons *J. Phys. Chem. A* **105** 9396-409
[28]    Zhan H, Zhang G, Zhang Y, Tan V B C, Bell J M and Gu Y 2016 Thermal conductivity of a new carbon nanotube analog: the diamond nanothread *Carbon* **98** 232-7
[29]    Stuart S J, Tutein A B and Harrison J A 2000 A reactive potential for hydrocarbons with intermolecular interactions *J. Chem. Phys.* **112** 6472-86
[30]    Zhan H, Zhang G, Bell J M and Gu Y 2016 The morphology and temperature dependent tensile properties of diamond nanothreads *Carbon* **In press**
[31]    Olbrich M, Mayer P and Trauner D 2015 Synthetic Studies toward Polytwistane Hydrocarbon Nanorods *The Journal of Organic Chemistry* **80** 2042-55
[32]    Jiang J-W, Wang J-S and Li B 2009 Thermal expansion in single-walled carbon nanotubes and graphene: Nonequilibrium Green's function approach *Phys. Rev. B* **80** 205429
[33]    Zhu L and Li B 2014 Low thermal conductivity in ultrathin carbon nanotube (2, 1) *Sci. Rep.* **4** 4917
[34]    Dickey J and Paskin A 1969 Computer simulation of the lattice dynamics of solids *Phys. Rev.* **188** 1407
[35]    Dickel D and Daw M S 2010 Improved calculation of vibrational mode lifetimes in anharmonic solids—Part I: Theory *Comput. Mater. Sci.* **47** 698-704
[36]    Gao Y, Wang H and Daw M 2015 Calculations of lattice vibrational mode lifetimes using Jazz: a Python wrapper for LAMMPS *Modell. Simul. Mater. Sci. Eng.* **23** 045002
[37]    Simkin M and Mahan G 2000 Minimum thermal conductivity of superlattices *Phys. Rev. Lett.* **84** 927
[38]    Chen Y, Li D, Lukes J R, Ni Z and Chen M 2005 Minimum superlattice thermal conductivity from molecular dynamics *Phys. Rev. B* **72** 174302
[39]    Zhu T and Ertekin E 2014 Phonon transport on two-dimensional graphene/boron nitride superlattices *Phys. Rev. B* **90** 195209
[40]    Latour B, Volz S and Chalopin Y 2014 Microscopic description of thermal-phonon coherence: From coherent transport to diffuse interface scattering in superlattices *Phys. Rev. B* **90** 014307
[41]    Chen G 1997 Size and interface effects on thermal conductivity of superlattices and periodic thin-film structures *J. Heat Transfer* **119** 220-9
[42]    Xiang B, Tsai C B, Lee C J, Yu D P and Chen Y Y 2006 Low-temperature specific heat of double wall carbon nanotubes *Solid State Commun.* **138** 516-20
[43]    Prasher R, Tong T and Majumdar A 2008 Approximate Analytical Models for Phonon Specific Heat and Ballistic Thermal Conductance of Nanowires *Nano Lett.* **8** 99-103
[44]    Padgett C W, Shenderova O and Brenner D W 2006 Thermal conductivity of diamond nanorods: molecular simulation and scaling relations *Nano Lett.* **6** 1827-31
[45]    Hu M and Poulikakos D 2012 Si/Ge Superlattice Nanowires with Ultralow Thermal Conductivity *Nano Lett.* **12** 5487-94
[46]    Schelling P K, Phillpot S R and Keblinski P 2002 Comparison of atomic-level simulation methods for computing thermal conductivity *Phys. Rev. B* **65** 144306

[47] Zhan H F, Zhang Y Y, Bell J M and Gu Y T 2014 Thermal conductivity of Si nanowires with faulted stacking layers *J. Phys. D: Appl. Phys.* **47** 015303
[48] Zhan H, Bell J M and Gu Y 2015 Carbon nanotube-based super nanotubes: tunable thermal conductivity in three dimensions *RSC Advances* **5** 48164-8
[49] Zhan H F, Zhang G, Bell J M and Gu Y T 2014 Thermal conductivity of configurable two-dimensional carbon nanotube architecture and strain modulation *Appl. Phys. Lett.* **105** 153105
[50] Liu X, Zhang G, Pei Q-X and Zhang Y-W 2013 Phonon thermal conductivity of monolayer MoS2 sheet and nanoribbons *Appl. Phys. Lett.* **103** 133113
[51] Zhan H, Zhang Y, Bell J M and Gu Y 2015 Suppressed Thermal Conductivity of Bilayer Graphene with Vacancy-Initiated Linkages *J. Phys. Chem. C* **119** 1748-52
[52] Zhu T and Ertekin E 2016 Phonons, Localization, and Thermal Conductivity of Diamond Nanothreads and Amorphous Graphene *Nano Lett.* **In press**
[53] Cao J X, Yan X H, Xiao Y and Ding J W 2004 Thermal conductivity of zigzag single-walled carbon nanotubes: Role of the umklapp process *Phys. Rev. B* **69** 073407
[54] Cao J X, Yan X H, Xiao Y, Tang Y and Ding J W 2003 Exact study of lattice dynamics of single-walled carbon nanotubes *Phys. Rev. B* **67** 045413
[55] Eucken A 1911 On the temperature dependence of the thermal conductivity of several gases *Phys. Z* **12** 1101-7
[56] Holland M 1964 Phonon scattering in semiconductors from thermal conductivity studies *Phys. Rev.* **134** A471
[57] Baughman R H 2006 Dangerously Seeking Linear Carbon *Science* **312** 1009-110
[58] Cannella C B and Goldman N 2015 Carbyne Fiber Synthesis within Evaporating Metallic Liquid Carbon *J. Phys. Chem. C* **119** 21605-11
[59] Robertson A W and Warner J H 2013 Atomic resolution imaging of graphene by transmission electron microscopy *Nanoscale* **5** 4079-93
[60] Liu X, Zhang G and Zhang Y-W 2015 Tunable Mechanical and Thermal Properties of One-Dimensional Carbyne Chain: Phase Transition and Microscopic Dynamics *J. Phys. Chem. C* **119** 24156-64
[61] Zhan H, Zhang G, Tan V B, Cheng Y, Bell J M, Zhang Y W, et al. 2016 Diamond nanothread as a new reinforcement for nanocomposites *Adv. Funct. Mater.* **In press**
[62] Meirovitch H 1983 Computer simulation of self‐avoiding walks: Testing the scanning method *J. Chem. Phys.* **79** 502-8
[63] Vasile C and Pascu M 2005 *Practical guide to polyethylene*: iSmithers Rapra Publishing)
[64] Sun H, Mumby S J, Maple J R and Hagler A T 1994 An ab Initio CFF93 All-Atom Force Field for Polycarbonates *J. Am. Chem. Soc.* **116** 2978-87
[65] Sun H 1998 COMPASS: an ab initio force-field optimized for condensed-phase applications overview with details on alkane and benzene compounds *J. Phys. Chem. B* **102** 7338-64
[66] Cooper C A, Cohen S R, Barber A H and Wagner H D 2002 Detachment of nanotubes from a polymer matrix *Appl. Phys. Lett.* **81** 3873-5
[67] Barber A H, Cohen S R and Wagner H D 2003 Measurement of carbon nanotube–polymer interfacial strength *Appl. Phys. Lett.* **82** 4140-2
[68] Al-Ostaz A, Pal G, Mantena P R and Cheng A 2008 Molecular dynamics simulation of SWCNT–polymer nanocomposite and its constituents *J. Mater. Sci.* **43** 164-73
[69] Guru K, Mishra S B and Shukla K K 2015 Effect of temperature and functionalization on the interfacial properties of CNT reinforced nanocomposites *Appl. Surf. Sci.* **349** 59-65
[70] Xiong Q L and Meguid S A 2015 Atomistic investigation of the interfacial mechanical characteristics of carbon nanotube reinforced epoxy composite *Eur. Polym. J.* **69** 1-15
[71] Liao K and Li S 2001 Interfacial characteristics of a carbon nanotube–polystyrene composite system *Appl. Phys. Lett.* **79** 4225-7
[72] Zheng Q, Xia D, Xue Q, Yan K, Gao X and Li Q 2009 Computational analysis of effect of modification on the interfacial characteristics of a carbon nanotube–polyethylene composite system *Appl. Surf. Sci.* **255** 3534-43
[73] Huang X, Qi X, Boey F and Zhang H 2012 Graphene-based composites *Chem. Soc. Rev.* **41** 666-86
[74] Wang Y, Zhan H, Xiang Y, Yang C, Wang C M and Zhang Y 2015 Effect of Covalent Functionalisation on Thermal Transport Across Graphene-Polymer Interfaces *J. Phys. Chem. C* **119** 12731-8
[75] Han Z and Fina A 2011 Thermal conductivity of carbon nanotubes and their polymer nanocomposites: A review *Prog. Polym. Sci.* **36** 914-44

[76]  Shen X, Wang Z, Wu Y, Liu X, He Y-B and Kim J-K 2016 Multilayer Graphene Enables Higher Efficiency in Improving Thermal Conductivities of Graphene/Epoxy Composites *Nano Lett.* **16** 3585-93
[77]  Dahl J, Liu S and Carlson R 2003 Isolation and structure of higher diamondoids, nanometer-sized diamond molecules *Science* **299** 96-9